\newcounter{myctr}
\def\myitem{\refstepcounter{myctr}\bibfont\noindent\ifnum\themyctr>9\else\phantom{0}\fi\hangindent17pt\themyctr.\enskip}

\documentclass{ws-ijqi}

\begin{document}

\markboth{Thiago Prud\^{e}ncio}
{Quantum state transfer and Hadamard gate for coherent states}

\catchline{}{}{}{}{}

\title{QUANTUM STATE TRANSFER AND HADAMARD GATE FOR COHERENT STATES}

\author{THIAGO PRUD\^ENCIO \thanks{Corresponding email: thprudencio@gmail.com
\vspace{6pt}}}

\address{Institute of Physics, University of Brasilia-UnB, 04455, 70919-970, Brasilia-DF, Brazil.}

\maketitle

\begin{abstract}
We propose a quantum state transfer from an 
atomic qubit to a cat-like qubit by means of one degenerate Raman interaction and one Hadamard gate operation 
for coherent states. We show that the coefficients of the atomic qubit can be 
mapped onto coherent state qubit, with an effective qubit state transfer. 
\end{abstract}


\section{INTRODUCTION}

Three-level atoms can appear in different configurations as $\Lambda$, $\Xi$ and V, what permits, for instance, the realization of
population inversion and many applications in laser physics \cite{walls,vedral,weisuter,sheldon}. 
By interacting with an one or two-mode optical field \cite{agarwal,nayak}, 
three-level atoms can in some special regimes be effectivelly described as two-level systems with adiabatic elimination of 
the highest level in the cases of $\Lambda$ \cite{wu} and $\Xi$ \cite{yang} configuration or the adiabatic elimination of
the lowest level in the case of V configuration \cite{xinhua}. In such situations, the system dynamics can be exactly 
solved for the Raman coupling with one or two modes of a quantized cavity electromagnetic field \cite{gerry}.

For a $\Lambda$ configuration, a special case occurs when the two lower levels are degenerate. In the interaction with a single-mode optical field 
at far off-resonant and large detuning regimes the atomic states are reduced to 
a two-level system of degenerate states, characterizing a degenerate Raman regime. This effective interaction leads in fact to 
an adequate description of the Raman interaction in far off-resonant and large detuning regimes if compared 
to the full microscopic Hamiltonian of the Raman process \cite{xuluo}, in the case of short evolving
times and description of physical quantities only involving the square of amplitude probabilities \cite{xu}.

From quantum optics, the degenerate Raman regime is well known, with important proposes of generation of non-classical states 
and protocols of quantum information
derived from it. For instance, the generation of Fock states \cite{avelarfilho,maiabaseia}, 
superpositions of coherent states \cite{zhengguo2}, superpositions of phase states \cite{avelarsouza} and 
generation of entangled coherent states \cite{song}. In quantum information, protocols of quantum teleportation
 were proposed for unknown atomic states \cite{zhengguo}, 
unknown entangled coherent states \cite{feng,song2} and superpositions of coherent states \cite{zheng,hassan}. 

However, a quantum information transfer from atom to field state have not yet been proposed 
for degenerate Raman regime. In fact, although the physics 
behind the atom-field entanglement is a well known process in this system \cite{raimond}, 
the step forward that can lead 
to the quantum transfer, i.e., a Hadamard gate for coherent state qubits, 
was only recently been proposed \cite{marek} and experimentally achieved \cite{tipsmark}. 

In this paper, we propose a quantum information transfer 
of an one-qubit state from an atom to a single mode field with a cavity initially in a coherent state. We first 
consider the atom-field interaction by degenerate Raman regime, then a Hadamard gate operation on the resulting coherent 
state qubit is realized, such that the state is lead to a qubit with the explict form of the atomic qubit. This propose 
has an advantage of realizing the quantum state transfer with a small number of operations. On the other hand, the requirement
of coherent state qubit, a cat like state, makes use of recent developments in the coherent state quantum 
computing \cite{cerf}. These states formed by superpositions of coherent states 
are the closest to classical superpositions due to the minimum uncertainty relation of coherent states,
related to the famous Schrodinger's example of cat states \cite{schrodinger}. Such states are, in fact, 
experimentally achievable, for instance, in electrodynamic cavities \cite{brune,raimond} and systems of trapped ions 
\cite{monroe}. The main experimental difficult for 
the creation and the observation of such superpositions of coherent states is related to the fast decay of 
the coherences \cite{huyet}. Advances in continuous variable quantum computation \cite{cerf,neegard} lead
to the manipulation of such states as qubit states and the possibility of realizing 
gate operations \cite{neegard}, both theoretical and experimentally \cite{tipsmark}. 

The paper is organized as follows: in the section II we review the 
degenerate Raman interaction between an atom and a coherent state. In 
the section III we discuss the Hadamard gate operation and the corresponding Hadamard gate operation for coherent states
 to be used in our protocol. In the section IV, we propose the protocol quantum state transfer of 
coherent state qubits making using of the previous sections. Finally, in the section V, 
we present our conclusions.

\section{DEGENERATE RAMAN INTERACTION}

\begin{figure}[h]
\centering
\includegraphics[scale=0.4]{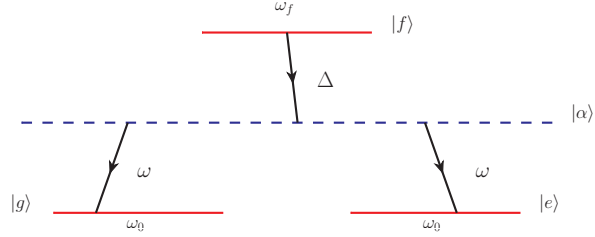}
\caption{(Color online) Scheme of degenerate Raman interaction.}
\label{gfr}
\end{figure}

We denote the three levels of the three-level Rydberg atom of $\Lambda$-type by
$|g\rangle$, $|e\rangle$ and $|f\rangle$. Due to the degeneracy, $|g\rangle$ and $|e\rangle$ 
have same frequency $\omega_{0}$. The frequency $\omega_{f}$ is associated to $|f\rangle$. 
The single-mode field is initially in a coherent state $|\alpha\rangle$ with frequency $\omega$ (See figure \ref{gfr}).
 Under atom-field degenerate Raman coupling the following relation is satisfied 
\begin{eqnarray}
\omega_{f}-\omega_{0} = \Delta + \omega,
\label{esg}
\end{eqnarray}
where $\Delta$ is the detuning between the atomic transition ($\omega_{f} -\omega_{0}$) and the single-mode frequency $\omega$. 
In the case of large detuning, the upper state $|f\rangle$ can be adiabatically eliminated \cite{gerry}.

In the case of large detuning, short evolving times and physical quantities that only 
involve the square of the amplitude probabilities, the following effective Hamiltonian can be considered ($\hbar=1$) \cite{xuluo}
\begin{eqnarray}
\hat{H}_{ef}= \hat{n}\beta(|e\rangle \langle g|+\rm{h.~c.}),
\label{ef1}
\end{eqnarray}
where $\beta = -\lambda^{2}/\Delta$ is the effective atom-field coupling, $\lambda$ is the transition
 coupling from the lower states ( $|e\rangle$ and $|g\rangle$) to the upper state ($|f\rangle$), $\hat{n}=\hat{a}^{\dagger}\hat{a}$ 
is the number operator,
$\hat{a}$ and $\hat{a}^{\dagger}$ the creation and annihilation operators acting on the single-mode field.

Retaining the Stark shifts in the adiabatic elimination\cite{sinatra} the hamiltonian term
\begin{eqnarray}
\hat{H}_{S}= \hat{n}\beta(|g\rangle \langle g|+|e\rangle \langle e|)
\end{eqnarray}
is included in the effective hamiltonian (\ref{ef1}), where
the Stark parameters are equal to the effective atom-field coupling $\beta$. We can then write 
the effective degenerate Raman hamiltonian as \cite{xu}
\begin{eqnarray}
\hat{H} = \hat{H}_{S} + \hat{H}_{ef}.
\label{efs}
\end{eqnarray}
For the single-mode field initially in the coherent state $|\alpha\rangle$, the hamiltonian (\ref{efs}) has validity when the following inequalities \cite{xu} are satisfied
\begin{eqnarray}
\Delta^{2} \gg 2|2\lambda\alpha|^{2},
\label{d}
\end{eqnarray}
and,
\begin{eqnarray}
t \ll \frac{3\Delta^{3}}{4|\lambda \alpha|^{4}}.
\label{b}
\end{eqnarray}

The time evolution during a time $t$ of an initial state of an atom in a superposed state of the form $c_{g}|g\rangle + c_{e}|e\rangle$,
$|c_{g}|^{2} + |c_{e}|^{2}=1$, and a field in a coherent state $|\alpha\rangle$ under the interaction (\ref{efs}) is given by 
\begin{eqnarray}
|\psi(t)\rangle &=& (c_{+}e^{-2i\hat{n}\beta t} -
c_{-})|g,\alpha\rangle \nonumber \\&+&(c_{+}e^{-2i\hat{n}\beta t} + c_{-})|e,\alpha\rangle,
\label{c}
\end{eqnarray}
where
\begin{eqnarray}
c_{\pm}=\frac{1}{2}\left(c_{e} \pm c_{g}\right).
\label{cpm}
\end{eqnarray}
We can also write the state (\ref{c}) as
\begin{eqnarray}
|\psi(t)\rangle &=& \left(c_{+}|e^{-2i\beta t}\alpha\rangle -c_{-}|\alpha\rangle \right)|g\rangle \nonumber \\
&+& \left(c_{+}|e^{-2i\beta t}\alpha\rangle +c_{-}|\alpha\rangle \right)|e\rangle.
\label{23e}
\end{eqnarray}
In the case $c_{g}=1$, $c_{e}=0$, corresponding to evolution of the state $|g\rangle$, we have
\begin{eqnarray}
|\psi(t)\rangle &=& \frac{1}{2}\left(|e^{-2i\beta t}\alpha\rangle +|\alpha\rangle \right)|g\rangle \nonumber\\
&+& \frac{1}{2}\left(|e^{-2i\beta t}\alpha\rangle -|\alpha\rangle \right)|e\rangle,
\end{eqnarray}
and in the case $c_{g}=0$, $c_{e}=1$, corresponding to evolution of the state $|e\rangle$, we have
\begin{eqnarray}
|\psi(t)\rangle &=& \frac{1}{2}\left(|e^{-2i\beta t}\alpha\rangle -|\alpha\rangle \right)|g\rangle \nonumber \\
&+& \frac{1}{2}\left(|e^{-2i\beta t}\alpha\rangle +|\alpha\rangle \right)|e\rangle.
\end{eqnarray}

Notice that for this situation can also deal with 
approximatelly degenerate states, where the two frequency transitions associated to each state 
by the same optical field have similar transition
frequencies. Consequently, this scheme an exact degeneracy is not necessary, we can have 
$\omega_{0}^{e}\approx \omega_{0}^{g}\approx \omega_{0}$, such that deviations can also be
easily incorporated. Since a small energy difference between the two lower atomic
states is likely to be typical in experiments, this situation where the states are approximatelly degenerates is 
of practical purpose.

\section{HADAMARD GATE FOR COHERENT STATE QUBITS}

A Hadamard gate operation is one case of one-gate operations and can be represented in matrix form by
\begin{eqnarray}
\hat{h}=\frac{1}{\sqrt{2}}\left( 
\begin{array}{cc}
1 &  \texttt{ }1 \\ 
1 & -1
\end{array}%
\right).
\end{eqnarray}
The action on states $|0\rangle$ and $|1\rangle$ leads respectivelly
\begin{eqnarray}
\hat{h}|0\rangle&=& \frac{|0\rangle + |1\rangle}{\sqrt{2}}\\
\hat{h}|1\rangle&=& \frac{|0\rangle - |1\rangle}{\sqrt{2}}
\end{eqnarray}
where $|0\rangle$ and $|1\rangle$ are orthonormal states.

In the case of coherent states a quantum gate operation can be realized only approximatelly due to the 
non-orthonormality of these states, with assintotically negligible overlap.

We consider a Hadamard gate operation for coherent states, involving coherent states qubits of the 
type $\left(a|\alpha\rangle +b|-\alpha\rangle \right)$. This is a particular one-qubit 
gate operation involving coherent states descring the one case of following general process
\begin{eqnarray}
\left(a|\alpha\rangle +b|-\alpha\rangle \right) \rightarrow \left(c|\alpha\rangle +d|-\alpha\rangle \right),
\end{eqnarray}
where $a$,$b$,$c$, $d$ are complex numbers. 

Note that taking into account the projection relations for coherent states \cite{walls}, in the case of the 
states $|\alpha\rangle$ 
and $|-\alpha\rangle$ ,  
\begin{eqnarray}
\langle \alpha | \alpha\rangle &=& 1, \label{c1}\\
\langle \alpha | -\alpha\rangle &=& e^{-2|\alpha|^{2}} \label{c2},
\end{eqnarray}
we can consider $|\alpha|$ suficiently large and satisfying 
degenerate Raman regime inequalities (\ref{b}) and (\ref{d}), such that the states $|\alpha\rangle$ 
and $|-\alpha\rangle$ are approximatelly orthonormal $\langle \alpha | -\alpha\rangle \approx 0$,    
a condition of negligible overlap. Consequently, the system of coherent states can be described 
as a Hilbert space formed by the states $|\alpha\rangle$ and $|-\alpha\rangle$.

A Hadamard gate operation can be represented by an operator $\hat{h}_{c}$ in the Hilbert space $\mathcal{H}_{c}$ formed by the 
states $|\alpha\rangle$ and $|-\alpha\rangle$. Its action on the qubit $\left(a|\alpha\rangle +b|-\alpha\rangle \right)$ 
will result in a change in the coefficients of the superposition
\begin{eqnarray}
\hat{h}_{c}\left(a|\alpha\rangle +b|-\alpha\rangle \right) \equiv \left(c|\alpha\rangle +d|-\alpha\rangle \right).
\end{eqnarray}
Different from qubits formed from the atomic states, $a|e\rangle + b|g\rangle$, or Fock states, 
$a|1\rangle + b|0\rangle$, for example, the qubit $a|\alpha\rangle +b|-\alpha\rangle$ is not physically simple of being 
operated. Indeed, these states are cat-like states and the gate operations in these states are a case 
of cat-like gate operations. 

These cat state qubits of coherent states, due the uncertainties between amplitude and phase are the minimum 
allowed by the Heisenberg's uncertainty principle in the case of coherent states \cite{walls}. Experiments for generation 
of superpositions of coherent states were proposed experimentally in both cavity QED \cite{brune} and trapped ions 
\cite{monroe}. The difficulty for 
the creation and the observation of such superpositions is mainly due to the fast decay of 
the coherences \cite{huyet}. With the advances of the techniques 
for continuous variable quantum computation, the necessity of deal with such states as qubit states leads to the 
problem of one and more gate operations \cite{neegard}. The realization of 
gate operations in superpositions of coherent states was only recently achieved \cite{tipsmark}, what rises the possibilities 
of quantum gate operations on qubits of cat-like states and use them in quantum computation.

In our case, the quantum gate operation on cat-like qubit will be of type
\begin{eqnarray}
\left(c_{+}|-\alpha\rangle + c_{-}|\alpha\rangle\right) \rightarrow \left( c_{g}|-\alpha\rangle +c_{e}|\alpha\rangle \right),
\end{eqnarray}
as we will show in the next section, this is a Hadamard type operation $\hat{h}_{c}$ that can be written in the basis of 
coherent states $|\alpha\rangle$ and $|-\alpha\rangle$ in the form
\begin{eqnarray}
\hat{h}_{c} &=&\frac{|-\alpha \rangle\langle -\alpha|-|\alpha \rangle\langle \alpha| 
+ |\alpha\rangle\langle -\alpha| + |-\alpha\rangle\langle \alpha|}{\sqrt{2}}.
\end{eqnarray}
Due to negligible overlap, we can map the coherent state qubit in a qubit space and the Hadamard gate 
can be represented in matrix form
\begin{eqnarray}
\hat{h}_{c}&\leftrightarrow&\frac{1}{\sqrt{2}}\left( 
\begin{array}{cc}
1 &  \texttt{ }1 \\ 
1 & -1
\end{array}%
\right), \\
|-\alpha\rangle &\leftrightarrow& \left( 
\begin{array}{c}
1  \\ 
0 
\end{array}%
\right), |\alpha\rangle \leftrightarrow \left( 
\begin{array}{c}
0  \\ 
1 
\end{array}%
\right).
\end{eqnarray}

\section{QUANTUM STATE TRANFER WITH COHERENT STATE QUBITS}

\begin{figure}[h]
\centering
\includegraphics[scale=0.4]{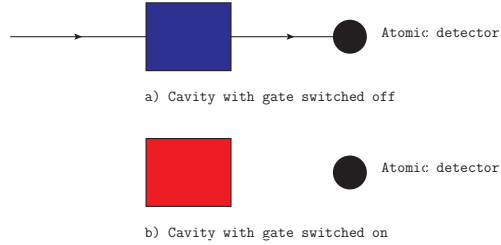}
\caption{(Color online) Protocol scheme divided in two steps.}
\label{gfu}
\end{figure}

Now we propose a quantum state transfer protocol as as coming from two steps (figure \ref{gfu}): 
(1) In the first step, the degenerate Raman interaction 
and an atomic detection are realized. In this step, we can consider 
the Hadamard gate for coherent states operations as switched off. 
(2) After the first step, 
we can switch on the gate operation for coherent states such that the process related to the manipulation of the 
coherent state qubit is achieved.

Let us start with the single-mode field initially in the coherent state $|\alpha\rangle$. 
At degenerate Raman interaction regime, the upper level $|f\rangle$ of the three 
level Rydberg atom of $\Lambda$-type can be neglected, reducing the atomic system to 
a two-level state system. In this situation, the atomic qubit state is described by
\begin{eqnarray}
|\phi\rangle= c_{g}|g\rangle + c_{e}| e\rangle,
\label{atls}
\end{eqnarray}
where
\begin{eqnarray}
|c_{g}|^{2} + |c_{e}|^{2}=1.
\end{eqnarray}
The interaction between the single-mode field in the coherent state $|\alpha\rangle$ and the atomic state $|\phi\rangle$, 
eq.(\ref{atls}), is given by the effective Raman interaction described previoulsly, eq. (\ref{efs}).

We want to realize the transfer of the unknown coefficients $c_{g}$ and $c_{e}$ from the atom 
to the single-mode field in such a way that in the final step the atomic qubit is lead to a coherent 
state qubit with the same form of the atomic one, i.e., carrying the coefficients $c_{g}$ and $c_{e}$. 

If the atom-field interaction occurs during a time $t$, the system composed of atom-field evolves to the state
\begin{eqnarray}
|\psi_{\alpha\alpha'}\rangle &=& \left(c_{+}|\alpha'\rangle -c_{-}|\alpha\rangle \right)|g\rangle \nonumber \\
&+& \left(c_{+}|\alpha'\rangle +c_{-}|\alpha\rangle \right)|e\rangle,
\end{eqnarray}
where $c_{\pm}$ is given by (\ref{cpm}) and $\alpha'$ is related to $\alpha$ by means of
\begin{eqnarray}
\alpha'&=& e^{-2i\beta t}\alpha.
\label{time}
\end{eqnarray}
An atomic detection then will project the field in a superposition of coherent states, leaving it in a cat-like state.
If the measurement of the atom is realized into the state $|e\rangle$, the field is projected into the cat state
\begin{eqnarray}
|\phi_{+}\rangle = c_{+}|\alpha'\rangle +c_{-}|\alpha\rangle.
\end{eqnarray}
On the other hand, a measurement of the atom in the state $|g\rangle$ will project the field into the cat state
\begin{eqnarray}
|\phi_{-}\rangle = c_{+}|\alpha'\rangle - c_{-}|\alpha\rangle.
\end{eqnarray}
We can choose the time of interaction $t=\pi/2\beta$ such that from (\ref{time}) we have
\begin{eqnarray}
|\phi_{\pm}\rangle = c_{+}|-\alpha\rangle \pm c_{-}|\alpha\rangle.
\label{finacod}
\end{eqnarray}
Now, taking into account the projection relations for coherent states, equations (\ref{c1}) and (\ref{c2}), 
and the condition of negligible overlap, $\langle \alpha | -\alpha\rangle \approx 0$, the states 
$|\alpha\rangle$ and $|-\alpha\rangle$ are orthonormal and can be described as a Hilbert space for a 
two-level system, generated by the 
coherent states $|\alpha\rangle$ and $|-\alpha\rangle$. In this situation, the atomic qubit can 
be stored as a coherent state qubit.  and 
capable of storing the qubit state expressed by the atomic state (\ref{atls}).

We note that the coeficients in (\ref{finacod}) do not correspond to $c_{g}$ and $c_{e}$, as in the initial atomic qubit, 
but are related to these by means of relations in (\ref{cpm}) or the following
\begin{eqnarray}
c_{+} + c_{-} &=& c_{e}, \\
c_{+} - c_{-} &=& c_{g}.
\end{eqnarray}
For this reason, it is necessary the realization of an qubit gate operation for coherent states. Indeed, a Hadamard gate operation 
can be written in the basis of coherent states by means of
\begin{eqnarray}
\hat{h}_{c}&=&|-\alpha \rangle\langle -\alpha|-|\alpha \rangle\langle \alpha| 
+ |\alpha\rangle\langle -\alpha| + |-\alpha\rangle\langle \alpha|,
\label{had}
\end{eqnarray}
where we drop the previous $\sqrt{2}$ factor and consider just the linear action on the states. As a $2\times2$-matrix, 
this operation can be written in terms of Pauli matrices as a linear combination 
of $\sigma_{z}$ and $\sigma_{x}$, by means of $\hat{h}_{c}=\sigma_{z} + \sigma_{x}$ \cite{chuang}. 

\begin{figure}[h]
\centering
\includegraphics[scale=0.5]{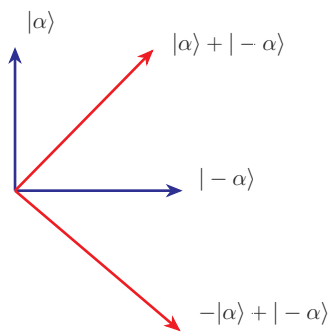}
\caption{(Color online) Scheme of the states 
$|\alpha\rangle$ and $|-\alpha\rangle$ states (blue arrows) and under action of the Hadamard gate for coherent states (red arrows).}
\label{gfrd}
\end{figure}

The quantum gate operation 
$\hat{h}_{c}$ acts on $|\alpha\rangle$ and $|-\alpha\rangle$ in the following way (see figure \ref{gfrd})
\begin{eqnarray}
\hat{h}_{c}|\alpha\rangle= -|\alpha\rangle + |-\alpha\rangle,
\end{eqnarray}
\begin{eqnarray}
\hat{h}_{c}|-\alpha\rangle= |\alpha\rangle + |-\alpha\rangle,
\end{eqnarray}

\begin{figure}[h]
\centering
\includegraphics[scale=0.5]{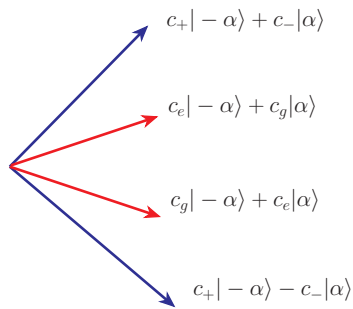}
\caption{(Color online) Scheme of $|\phi_{+}\rangle$ and $|\phi_{-}\rangle$ states (blue arrows) 
and under action of the Hadamard gate for coherent states (red arrows).}
\label{gfrd23}
\end{figure}

The action of the operator $\hat{h}_{c}$ on the field $|\phi_{+}\rangle$ leads the single mode field to the following state 
\begin{eqnarray}
\hat{h}_{c}|\phi_{+}\rangle= c_{e}|-\alpha\rangle +c_{g}|\alpha\rangle,
\label{dasxz}
\end{eqnarray}
corresponding to a quantum information transfer when the atom is detected in the state $|e\rangle$.
On the other hand, the action of $\hat{h}_{c}$ on the field $|\phi_{-}\rangle$ leads to the state 
\begin{eqnarray}
\hat{h}_{c}|\phi_{-}\rangle = c_{g}|-\alpha\rangle +c_{e}|\alpha\rangle,
\label{dasxz2}
\end{eqnarray}
corresponding to a quantum information transfer when the atom is detected in the state $|g\rangle$ (see figure \ref{gfrd23}).

In order to fix the coefficients, we can choose 
the previous selective atomic state detection to measure the atomic state in the ground state $|g\rangle$, 
such that the coherent state qubit has the form (\ref{dasxz2}). Consequently, this protocol is equivalent 
to a quantum state transfer of 
the state $c_{g}|0\rangle + c_{e}|1\rangle$, that in our case goes from an atomic qubit state for a coherent state qubit, 
a cat-like qubit state. 

\section{CONCLUSION}

In conclusion, a quantum information transfer from atom to field state had not yet been proposed 
for degenerate Raman regime. In fact, although the physics behind the atom-field entanglement it is a well known 
process \cite{suter}, the step forward of a Hadamard gate for coherent state qubits 
was only recently proposed \cite{marek} and experimentally achieved \cite{tipsmark}, leading to the possibility of realize  
quantum computation with cat-like states.

Here we have proposed a protocol of quantum transfer 
where the atom-field interacts in degenerate Raman regime, 
a selective atomic detection of one of the degenerate states is realized 
and finally a Hadamard gate operation for coherent states is applied in the coherent state qubit. 
This protocol is equivalent to a quantum state transfer of 
the state $c_{g}|0\rangle + c_{e}|1\rangle$, that in our case goes from an atomic qubit state for a coherent state qubit, 
a cat-like qubit state. As such, it can make use of the recent advances in coherent state quantum computation. 

This scheme can also take into account approximatelly degenerate states, such that
the two transitions frequences associated to each state by the same optical field have similar transition
frequencies. Consequently, an exact degeneracy is not necessary. Since a small energy difference between the two lower atomic
states is likely to be typical in experiments, this is of practical interest.

Independent of the final state in which the atom is detected, $|g\rangle$ or $|e\rangle$, 
the action of the Hadamard gate for coherent states $\hat{h}_{c}$ on the resulting coherent state qubit 
leads to a complete quantum transfer of the atomic qubit coefficients $c_{g}$ and $c_{e}$ 
from the atom to the single-mode field. However, in order to avoid the
 statistical average that traces over the atom state resulting in a 
statistical mixture of the results on detection in the states $|g\rangle$ and $|e\rangle$, we have 
to choose one state to realize the detection. By choosing the atomic selective detection on 
the state $|g\rangle$, we can complete the quantum state transfer of 
the state $c_{g}|0\rangle + c_{e}|1\rangle$, from the atomic qubit to the coherent state qubit.

The experimental implementation of the atom-field entanglement is generally considered with the use
of Rydberg atoms of long lifetime and high-$Q$ superconducting microwave cavities, where 
the dissipative processes are not
relevant compared to the the atom-field interaction \cite{song}. The appropriate 
choices of $\lambda$ and $\Delta$, of the order $\approx 10$ kHz and $\approx 10^{3}$ kHz, respectivelly,
are sufficient for the condition of large detuning. A quality factor $Q$ of order 
$\approx 10^{11}$ corresponds to a cavity lifetime $T_{c}$ of the order $\approx 10^{-1}$ s, associated to atomic velocities 
of order $\approx 10^{3}$ m/s \cite{zoller}. We can choose $\alpha$ equal to $10$, $5$, $3$ or $2$, for instance, that 
lead to negligible overlap conditions $e^{-2|\alpha|^{2}}$ of orders $\approx 10^{-87}$, $\approx 10^{-22}$, $\approx 10^{-8}$ and 
$\approx 10^{-4}$, respectivelly. The Hadamard gate operation for coherent states can be executed in $\approx 10^{-2}$ s without surpass the cavity lifetime \cite{tipsmark}.

The quantum transfer from atom qubit to cat-like state qubit is an 
important quantum operation for coherent state quantum computing \cite{neegard,cerf,cochrane,jeong,ralph,lund}. 
When the coefficients of the atomic qubit 
are mapped onto coherent state qubit, the cat-like state, a quantum state transfer between
systems of different nature is realized with the effective transfer of the qubit $c_{g}|0\rangle + c_{e}|1\rangle$. This is 
an interesting instance of matter-light quantum state transfer that has recently been achieved with different protocols in
quantum information and quantum computation \cite{raimond,zheng4,wang,hagley,osnaghi,davidovich,matsukevich,sherson,koshino,maitre,barz}.

\section*{ACKNOWLEDGEMENTS}
The author thanks CAPES (Brazil) for partial finantial support and
reviewers for suggestions of references and comments on the manuscript.

\end{document}